# Dielectric Engineering of HfO$_2$ Gate Stacks Towards Normally-ON and Normally-OFF GaN HEMTs on Silicon

Hareesh Chandrasekar, Sandeep Kumar, K. L. Ganapathi, Shreesha Prabhu, Surani Bin Dolmanan, Sudhiranjan Tripathy, Srinivasan Raghavan, K. N. Bhat, *Member, IEEE,* Sangeneni Mohan, R. Muralidharan, *Member, IEEE,* Navakanta Bhat, *Senior Member, IEEE* and Digbijoy N. Nath

*Abstract*— We report on the interfacial electronic properties of HfO$_2$ gate dielectrics both, with GaN towards normally-OFF recessed HEMT architectures and the AlGaN barrier for normally-ON AlGaN/GaN MISHEMTs for GaN device platforms on Si. A conduction band offset of 1.9 eV is extracted for HfO$_2$/GaN along with a very low density of fixed bulk and interfacial charges. Conductance measurements on HfO$_2$/GaN MOSCAPs reveal an interface trap state continuum with a density of 9.37x10$^{12}$ eV$^{-1}$cm$^{-2}$ centered at 0.48 eV below E$_C$. The forward and reverse current densities are shown to be governed by Fowler-Nordheim tunneling and Poole-Frenkel emission respectively. Normally-ON HfO$_2$/AlGaN/GaN MISHEMTs exhibit negligible shifts in threshold voltage, transconductances of 110mS/mm for 3 µm gate length devices, and three-terminal OFF-state gate leakage currents of 20 nA/mm at a V$_D$ of 100 V. Dynamic capacitance dispersion measurements show two peaks at the AlGaN/GaN interface corresponding to slow and fast interface traps with a peak D$_{it}$ of 5.5x10$^{13}$ eV$^{-1}$cm$^{-2}$ and 1.5x10$^{13}$ eV$^{-1}$cm$^{-2}$ at trap levels 0.55 eV and 0.46 eV below E$_C$ respectively. The HfO$_2$/AlGaN interface exhibits a peak D$_{it}$ of 4.4x10$^{13}$ eV$^{-1}$cm$^{-2}$ at 0.45 eV below E$_C$.

## I. INTRODUCTION

High-k gate dielectrics are being increasingly adopted for both normally-ON and normally-OFF AlGaN/GaN transistor architectures as they afford superior control over gate leakage and channel electrostatics. However, the presence of such a dielectric layer adds another interface, and therefore another layer of complexity, to a material system whose interfacial properties play a key role in device operation. For instance, typically employed Al$_2$O$_3$ dielectrics have shown a very high positive fixed charge density of 4.6x10$^{12}$ cm$^{-2}$,[1] which results in significant shifts in threshold voltage of the fabricated devices, in addition to impacting reliable operation.

Given the sensitivity of device performance to gate oxide deposition and processing conditions, dielectric engineering is crucial for controlling the electrostatics of GaN transistors.

A variety of gate dielectrics have been investigated for the AlGaN/GaN material system ranging from silicon nitride and silicon oxide to Al$_2$O$_3$ and HfO$_2$, and multi-layers gate stacks.[1-7] Among them, HfO$_2$ is promising in view of its high dielectric constant (>19), large band gap (5.8 eV), low leakage and its compatibility with CMOS processing, which is of particular relevance for the technologically and commercially important GaN-on-Si device platform co-processed in existing silicon foundry lines. While device demonstrations of AlGaN/GaN HEMTs using HfO$_2$ gate dielectrics with good performance metrics have been reported [8-15], reports on the interfacial electronic properties of such HfO$_2$/GaN and HfO$_2$/AlGaN/GaN interfaces for normally-OFF and normally-ON operation are limited.[16] A thorough investigation of the same is expected to be timely and important to develop appropriate device designs and processing strategies to maximize the performance of HfO$_2$ based dielectric schemes.

Here we report on the use of e-beam evaporated HfO$_2$ as high-k gate dielectrics for GaN-on-Si device platforms. Extensive electrical characterization was carried out to determine the interfacial properties of HfO$_2$ with both, GaN directly, towards normally-OFF device architectures, and with AlGaN barriers for normally-ON AlGaN/GaN MISHEMT structures. Band offsets and fixed oxide charge densities, interface trap densities and trap time constants with their energy distribution through the band gap have been estimated. Gate leakage measurements and MISHEMT characterization were also performed to evaluate the effect of e-beam evaporated HfO$_2$ films on device operation. These results are expected to inform the development of further processing schemes to control and characterize interfacial traps at HfO$_2$/(Al)GaN interfaces.

## II. EXPERIMENTAL DETAILS

Si-doped n-type GaN films were grown on Si substrates by MOCVD using an AlN nucleation layer, step-graded AlGaN transition layers as reported elsewhere.[17, 18], with the electron concentration estimated to be 7x10$^{16}$ cm$^{-3}$ by both Hall and C-V measurements.[18] Al electrodes were used as

The work is funded by Ministry of Electronics and IT (MeitY) under its NAMPET project. H. Chandrasekar, S. Kumar, K.L. Ganapathi, S. Prabhu, S. Raghavan, K.N. Bhat, S. Mohan, R. Muralidharan, N. Bhat and D.N. Nath are with the Centre for Nano Science and Engineering, Indian Institute of Science, CV Raman Road, Bangalore 560012, India (e-mail: hareeshc2408@gmail.com; digbijoy@cense.iisc.ernet.in).

S.B. Dolmanan and S. Tripathy are with the Institute of Materials Research and Engineering (IMRE), Agency for Science, Technology, and Research (A*STAR), Singapore.

ohmic contacts. HfO$_2$ films of different thicknesses - 10, 20 and 30 nm - were deposited using e-beam evaporation after a HCl pre-treatment, followed by post-deposition annealing at 300°C for 5 min in a forming gas ambient.[19] Ni/Au films were then patterned as gate electrodes and a post-metallization anneal at 350°C for 5 min in a forming gas ambient completed fabrication of the MOSCAP structures. A wide variety of PDA and PMA conditions were explored in RTA and the optimal conditions were chosen so as to minimize hysteresis and maximize the dielectric constant as extracted from C-V measurements, while keeping the annealing temperatures low to minimize interfacial reactions.

For fabrication of MISHEMTs, a HEMT stack of ~3.6 – 3.7 μm was grown on 1 mm thick Si (111) substrates of 200 mm diameter using MOCVD. The growth stack consisted of an AlN nucleation layer followed by step-graded AlGaN transition layers and ~1.0 μm thick C-doped buffer. The top 2DEG layers consist of 300 nm undoped GaN, 1 nm AlN spacer, 16 nm Al$_{0.27}$Ga$_{0.73}$N barrier layer and 2.0 nm thin GaN cap. Ti/Al/Ni/Au ohmic deposition and RTA was followed by mesa etching in a BCl$_3$/Cl$_2$ ICP-RIE. E-beam evaporated 30 nm HfO$_2$ films were deposited as the gate dielectric followed by Ni/Au gate electrodes using the optimized PDA/PMA conditions determined from the MOSCAP studies. PECVD SiN$_x$ of 1 μm was used for passivation. Circular capacitors of were used for C-V and conductance measurements, while the transistors reported in this study have a L$_{gs}$, L$_{gd}$ and L$_g$ of 2, 4 and 3 μm respectively. Electrical properties of the HfO$_2$/n-GaN MOSCAP and HfO$_2$/AlGaN/GaN MISHEMT structures including C-V, conductance, I-V, and transistor characteristics were measured using an Agilent B1500 semiconductor device analyzer equipped with a capacitance measurement module.

## III. RESULTS AND DISCUSSION

### A. HfO$_2$/GaN MOSCAP Interfacial Characterization

In order to estimate the interfacial charge density and band offsets between HfO$_2$ and GaN, three different HfO$_2$ thicknesses - 10, 20 and 30 nm - were used for C-V analysis. Fig. 1 shows the 1 MHz capacitance-voltage curves for these three dielectric thickness, normalized with respect to the area. Typically, the presence of a finite interfacial charge density leads to an electric field drop across the dielectric whose voltage is thickness dependent, giving rise to a flatband voltage (V$_{FB}$) shift with changing dielectric thickness. The slope of V$_{FB}$ versus dielectric thickness plot can then be used to estimate electric field drop and hence the interfacial charge from Poisson's equation, and the conduction band offset between the GaN and gate dielectric as given in (1),[1]

$$qV_g = -q\mathrm{E}_{ox}t_{ox} + \varphi_b - \Delta E_C - \varphi_s \quad (1)$$

where φ$_b$ is the barrier height between the gate metal and the dielectric, φ$_s$, the location of Fermi level in GaN with respect to E$_C$, ΔE$_C$ the conduction band offset at the dielectric-GaN interface, E$_{ox}$ and t$_{ox}$ the electric field across the dielectric and dielectric thickness respectively. Positive slopes of V$_{FB}$ vs t$_{ox}$ correspond to negative interfacial charge densities and vice-versa and are plotted in Fig. 2 for the fabricated HfO$_2$/GaN MOSCAPs.

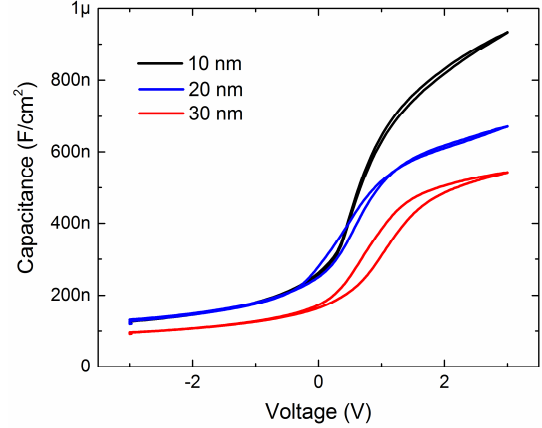

Fig. 1. Normalized capacitance-voltage curves of 10, 20 and 30 nm thick e-beam evaporated HfO$_2$ films with optimized post-deposition and post-metallization annealing at 1 MHz measurement frequency. The annealing conditions were optimized to obtain maximum dielectric constant with minimal hysteresis in the CV.

We see that the variation of V$_{FB}$ with dielectric thickness for the optimally annealed samples is insignificant and well within the experimental error corresponding to device-to-device variability. In comparison, ALD Al$_2$O$_3$ layers on GaN have been shown to possess a very high positive interfacial charge density of 4.6x10$^{12}$ cm$^{-2}$.[1] Such low interfacial charge densities on HfO$_2$ gates would lead to lower threshold voltage shifts post-device fabrication and more reliable operation. Using the Ni/HfO$_2$ barrier height of 2.4 eV,[20] and given that E$_F$ is 91 meV below the conduction band for n-GaN of 7x10$^{16}$ cm$^{-3}$ doping density, we estimate a conduction band offset between HfO$_2$ and GaN of 1.9 eV. This value compares well with prior reports on HfO$_2$/GaN conduction band offsets of 2.1±0.1 eV measured using XPS.[21] The combination of such a reasonably high band offset with GaN and the low interfacial charge density makes HfO$_2$ an attractive gate dielectric for normally-OFF recessed HEMT architectures. On the other hand, HfO$_2$/GaN samples processed under non-optimal (higher temperature) annealing conditions show a significant shift in V$_{FB}$ with t$_{ox}$ as shown in Fig. 2 (right). This corresponds to an oxide field drop of 0.2 MV/cm and a high negative charge density of 2.1x10$^{12}$ cm$^{-2}$ at the interface. The extracted band offset between HfO$_2$/GaN also reduces to 1.4 eV strongly suggesting the possibility of interfacial reactions between the two layers. The difference between these two set of samples highlights the importance of a careful and informed choice of deposition and processing conditions to realize the full potential of HfO$_2$ as a gate stack.

Trap density at the HfO$_2$/GaN interface is the other key factor that determines the performance of any fabricated devices, and characterization of the traps is hence of utmost importance. Conductance measurements were carried out on the HfO$_2$/GaN MOSCAPs to determine trap densities, characteristic time constants and locations within the band gap. The measured conductance and capacitances were corrected for series resistance given by,[22, 23]

$$R_s = \frac{G_{m,a}}{G_{m,a}^2 + \omega^2 C_{m,a}^2} \quad (2)$$

where $G_{m,a}$ and $C_{m,a}$ are the measured accumulation conductance and capacitances and $\omega$, the angular frequency.

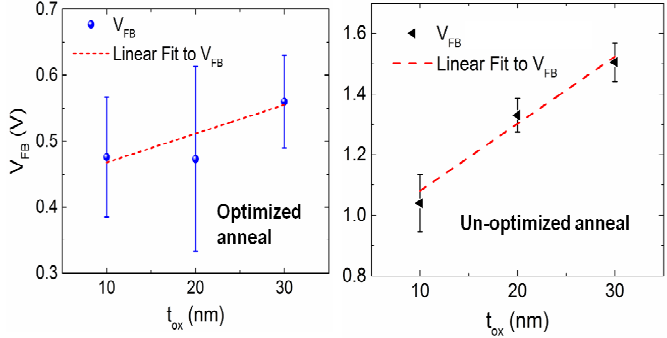

Fig. 2. Flatband voltage vs dielectric thickness plots of $HfO_2$/GaN MOSCAPs for the optimized (left) and the un-optimized (high temperature annealed) PDA/PMA conditions (right). The error bars indicate standard deviation over four measured devices for each dielectric thickness.

The capacitance and conductance were then corrected for this series resistance using (3) and (4).

$$C_c = \frac{(G_m^2 + \omega^2 C_m^2)C_m}{(G_m - R_s(G_m^2 + \omega^2 C_m^2))^2 + \omega^2 C_m^2} \quad (3)$$

$$G_c = \frac{(G_m^2 + \omega^2 C_m^2)[G_m - R_s(G_m^2 + \omega^2 C_m^2)]}{[G_m - R_s(G_m^2 + \omega^2 C_m^2)]^2 + \omega^2 C_m^2} \quad (4)$$

where the terms $G_m$, $C_m$, $R_s$ and $\omega$ have their usual meanings. The parallel conductance from interface traps is then calculated as,

$$G_p = \frac{\omega^2 C_{acc}^2 G_c}{G_c^2 + \omega^2 (C_{acc} - C_c)^2} \quad (5)$$

where $C_{acc}$ represents the accumulation capacitance, with $G_c$ and $C_c$ being the corrected conductance and capacitances. Assuming a continuum of trap states distributed in the band gap, the interface trap density is then estimated from a plot of $G_p/\omega$ vs $\omega$ as[24]

$$\frac{G_p}{\omega} = \frac{qD_{it}}{2\omega\tau_{it}} \ln[1 + (\omega\tau_{it})^2] \quad (6)$$

with $D_{it}$ and $\tau_{it}$ being the interface trap densities in $eV^{-1}cm^{-2}$ and the trap time constant in seconds.

A plot of the normalized $G_p/\omega$ as a function of measurement frequency, estimated from (2)-(5), is shown in Fig. 3 for a representative 10 nm $HfO_2$ MOSCAP and different gate voltages. Fitting these curves to (6) yields a peak $D_{it}$ of $9.37 \times 10^{12}$ $eV^{-1}cm^{-2}$ and a characteristic time constant of 1.48 μs. The position of the trap level below $E_c$ can be estimated as

$$E_C - E_T = kT \ln(\sigma_c v_{th} D_{DOS} \tau) \quad (7)$$

where $\sigma_c$ refers to the capture cross-section of the trap, $v_{th}$ the thermal velocity, $D_{DOS}$, the density of states in the conduction band for this case and $\tau$ is the trap response time. Assuming a nominal capture cross-section of $4 \times 10^{-13}$ $cm^{-2}$[25], $D_{DOS}$ for GaN as $2.2 \times 10^{18}$ $cm^{-3}$, we obtain a trap level 0.48eV below $E_C$.

These $D_{it}$ values are comparable to prior reports for ALD $Al_2O_3$ dielectrics deposited on GaN directly[26] and further indicate the promise of e-beam evaporated $HfO_2$ for normally-OFF recessed-gate device architectures.

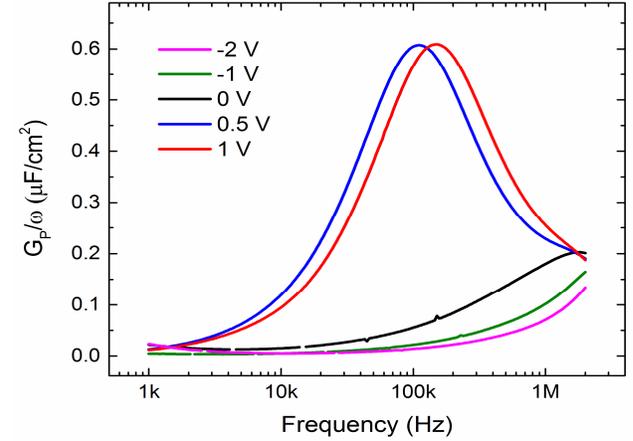

Fig. 3. Normalized $G_p/\omega$ vs measurement frequency curves of a representative 10nm $HfO_2$/GaN MOSCAP for various gate voltages. The peaks correspond to a $D_{it}$ of $9.37 \times 10^{12}$ $eV^{-1}cm^{-2}$ and a trap time constant of 1.48μs, centered at 0.48eV below $E_c$.

The other key factor influencing the choice of gate dielectrics is gate leakage. Fig. 4 shows the forward and reverse leakage current densities for the $HfO_2$/GaN samples. We see that the forward current densities are ~1A/$cm^2$ and thickness independent, while reverse currents vary from 1-200 μA/$cm^2$ ($V_G$ =-3 V) as the dielectric thickness varies from 10 to 30 nm. These numbers compare favorably with reported leakage current densities for ALD $HfO_2$ gate stacks, [27] which however showed a thickness dependence for forward currents and hence a different current transport mechanism from the e-beam evaporated films in the present study. Upon further analysis, forward current densities were found to follow a Fowler-Nordheim $\ln(J/E^2)$ vs $1/E$ dependence,[28]

$$J = C_{FN,1} E^2 \exp(-\frac{C_{FN,2}}{E}) \quad (7)$$

where $C_{FN,1}$ and $C_{FN,2}$ are constants and J and E refer to current densities and oxide electric field respectively. The inset at the bottom-right of Fig. 4 shows the F-N plot of the measured currents along with linear fits in the accumulation region for all three dielectric thicknesses. Similarly, reverse currents were found to be governed by Poole-Frenkel emission from trap states with a electric field dependence given by (8),[28]

$$J = C_{PF} E \exp[-\frac{q(\phi_B - \sqrt{qE/\pi\varepsilon_{ox}})}{kT}] \quad (8)$$

where $C_{PF}$ is the Poole-Frenkel constant, $\varphi_B$ the location of the trap level below $E_C$, $\varepsilon_{ox}$ the dielectric constant and J and E are the current densities and oxide electric fields. The inset at the top left of Fig. 4 shows the measured data and linear fits of $\ln(J/E)$ vs $\sqrt{E}$ for the three dielectric thicknesses in inversion, indicative of P-F transport through $HfO_2$ films. A detailed investigation of the leakage mechanisms in these $HfO_2$-GaN stacks is of topical interest and will be presented in a future work.

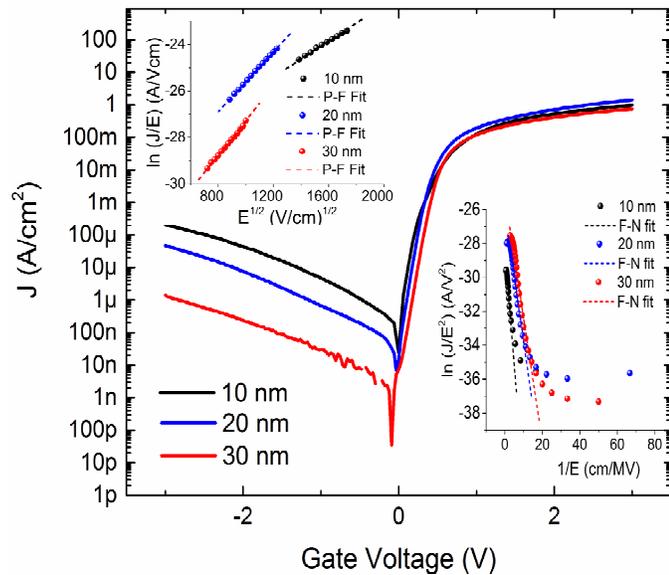

Fig. 4. Forward and reverse gate leakage current densities as a function of gate voltage for different $HfO_2$ thicknesses. Top inset: Poole-Frenkel transport determines reverse leakage currents as shown in the linear fits to the P-F characteristic plot of $\ln(J/E)$ vs $E^{1/2}$. Bottom inset: Fowler-Nordheim tunneling is responsible for the forward currents, seen from the straight line fits of F-N tunneling plot of $\ln(J/E^2)$ vs $1/E$.

### B. $HfO_2$/AlGaN/GaN MISHEMT Interfacial Characterization

To evaluate the device performance of e-beam evaporated $HfO_2$ gate stacks for normally-ON devices, $HfO_2$/AlGaN/GaN MISHEMTs with $HfO_2$ thickness of 30 nm were studied. The output and transfer curves of these devices can be seen in Fig. 5.

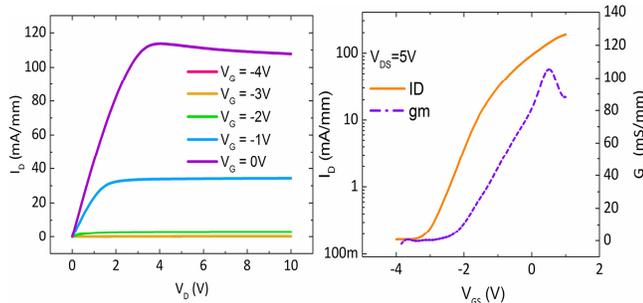

Fig. 5. Output (left) and transfer characteristics (right) of a representative 30nm-$HfO_2$/AlGaN/GaN MISHEMT. The threshold voltage is at -3.5V, same as for Schottky HEMTs on these samples, indicating the absence of interfacial charges. A peak transconductance of 110 mS/mm was observed for these samples.

The threshold voltages for the $HfO_2$ MISHEMTs were -3.5V which is almost identical to the threshold voltage observed for Schottky gate HEMTs fabricated using these samples. Drain current modulation with applied gate bias can be observed from the output characteristics of Fig. 5 (left) and transconductance values of 110 mS/mm were obtained for devices with a gate length of 3 μm. These values are similar to those (130 mS/mm) reported for 25 nm thick ALD $HfO_2$ MISHEMTs with the same gate length previously.[15] To estimate gate leakages in these stacks, drain and gate currents versus gate voltage and three-terminal OFF-state leakages for drain voltages up to 100 V are shown in Fig. 6. We see that the maximum gate leakage current is 600 nA/mm for a forward bias of 1 V on the gate. On the other hand, OFF-state gate leakages at gate bias of -3.5 V and -4.5 V and drain voltages of 100V are 15-20 nA/mm for these devices, representative of the excellent reverse blocking characteristics of e-beam evaporated $HfO_2$ gate stacks as discussed earlier in Fig. 4. These results further indicate the advantages of $HfO_2$ as gate dielectrics for normally-ON AlGaN/GaN HEMTs.

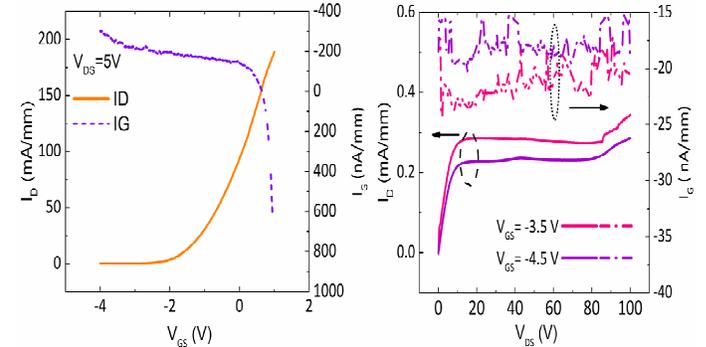

Fig. 6. (Left) Drain and gate leakage currents, normalized with device width, as a function of gate voltage at a fixed drain bias of 5V. A maximum gate leakage of 600nA/mm was observed for a forward bias of +1V. (Right) Three-terminal OFF-state drain and gate leakages for drain voltages up to 100V. Very low gate leakage currents of 15-20 nA/mm were observed for reverse blocking through $HfO_2$ gate stacks.

In order to determine the interfacial properties at the $HfO_2$/AlGaN/GaN interface, circular MISHEMT test structures were characterized. The conductance technique was used to determine the interfacial trap densities using (2)-(6), and their variations within the band gap was probed by changing the gate voltage. Biasing the structures in depletion close to their threshold voltages leads to trap response at the AlGaN/GaN interface corresponding to variations in the Fermi level with the applied a.c. excitation. The trap states at the $HfO_2$/AlGaN interface, on the other hand, respond to the structure being biased in accumulation when those at the AlGaN/GaN interface are well below the Fermi level and do not respond. Hence the trap states at both the AlGaN/GaN and $HfO_2$/AlGaN can be sequentially probed by varying the applied bias in the depletion and accumulation regions (also called the dynamic capacitance dispersion technique).[29]

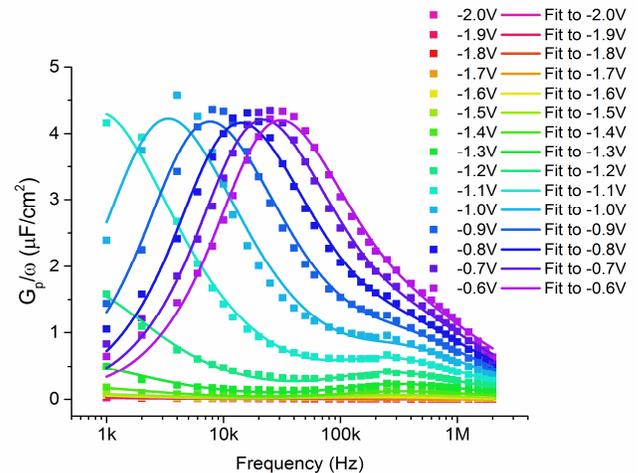

Fig. 7. Normalized $G_p/\omega$ vs measured frequency plots for normally-ON $HfO_2$/AlGaN/GaN MISHEMT capacitor structures with varying gate bias in depletion. Also shown are two peak fits to (9) corresponding to the presence of slow and fast trap states at the AlGaN/GaN interface.

The normalized parallel conductance plot in depletion as a function of frequency is shown in Fig. 7 for various depletion bias voltages. A good fit for these peaks was obtained using two-peak terms of (9), analogous to the single-trap contribution of (6).

$$\frac{G_p}{\omega} = \frac{qD_{it,1}}{2\omega\tau_{it,1}}\ln[1+(\omega\tau_{it,1})^2] + \frac{qD_{it,2}}{2\omega\tau_{it,2}}\ln[1+(\omega\tau_{it,2})^2] \quad (9)$$

The presence of two trap states as inferred from the fitting, and their respective variation in trap densities and characteristic time constants within the band gap were extracted using (9) and plotted in Fig. 8. These two trap states have been characterized as "slow" and "fast" for the purpose of discussion, with the slow traps having a response time ≥10μs and fast traps <10 μs. The slow trap state had a peak density of $5.5\times10^{13}$ $eV^{-1}cm^{-2}$ at 0.5 eV below the conduction band edge which reduces by two orders of magnitude at 0.62 eV below $E_C$. The trap time constant showed significant dispersion of 12 μs to 1.5 ms with applied bias, characteristic of efficient band bending at the AlGaN/GaN interface. Similarly, the density of fast traps varied from $1.5\times10^{13}$ $eV^{-1}cm^{-2}$ at 0.41 eV below the band edge and reduces to a very low $9\times10^{10}$ $eV^{-1}cm^{-2}$ at 0.5 eV below $E_C$, with large time constant dispersions from 10 μs to 400 ns, over applied gate bias again indicating effective band bending and a good quality AlGaN/GaN interface.[23]

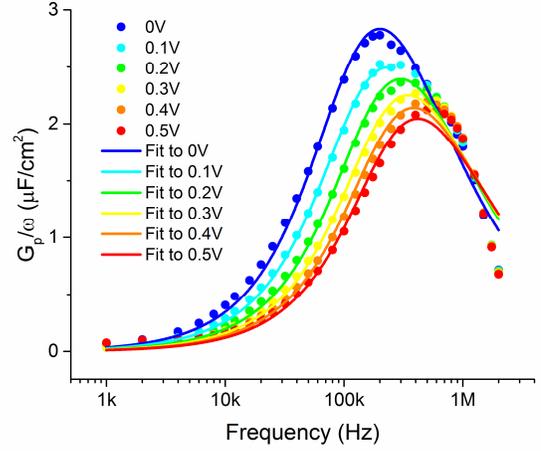

Fig. 9. Normalized $G_p/\omega$ vs measured frequency plots for normally-ON $HfO_2$/AlGaN/GaN MISHEMT structures with varying gate bias in accumulation. Also shown are peak fits to (6) corresponding to the interface trap response at the $HfO_2$/AlGaN interface.

Even though the samples were subject to an HCl pre-treatment for native oxide removal and low temperature annealing steps ensure the prevention of interfacial layer formation at this interface, additional pre-treatments seem to be warranted in order to de-pin the Fermi level at the $HfO_2$/AlGaN interface. However, the effect of such Fermi level pinning on device performance, given the higher near band edge $D_{it}$s at the AlGaN/GaN interface, is not expected to be significant.

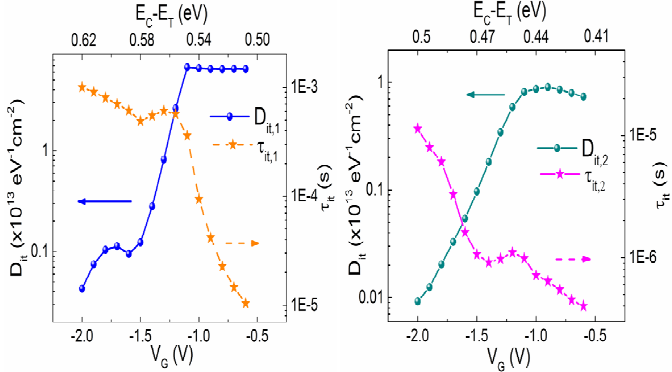

Fig. 8. Variation of interface trap densities and characteristic trap time constants at the AlGaN/GaN interface for slow (left) and fast traps (right) with applied gate voltage/energy level within the band gap.

Furthermore, normalized conductance versus frequency plots for the accumulation region and interface trap density fits corresponding to trap state response at the $HfO_2$/AlGaN interface, are shown in Fig. 9. A single interface trap state continuum was found to contribute to trap conductance at this interface and the trap parameters were extracted from fits to (6), also shown in Fig. 9. The evolution of interface state densities and trap response times with applied bias within the AlGaN bandgap are summarized in Fig. 10. The $HfO_2$/AlGaN interface had a peak $D_{it}$ of $4.4\times10^{13}$ $eV^{-1}cm^{-2}$ at 0.45eV below $E_C$ and shows minimal variation in time response times with gate voltage - varying from 0.75 μs to 1.56 μs. While this value of trap density is comparable to reported values for $Al_2O_3$/AlGaN interfaces,[29] the lack of variation of trap response time with applied bias indicates that the Fermi level is pinned at $HfO_2$/AlGaN interface similar to the $HfO_2$/GaN interface.

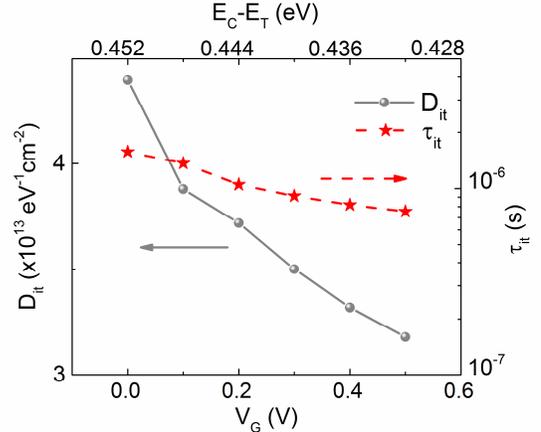

Fig. 10. Interface trap densities and trap response times as a function of applied voltage/energy level within the AlGaN band gap. A peak $D_{it}$ of 4.4e13 $eV^{-1}cm^{-2}$ was observed at a trap level of 0.45eV below the band gap.

## IV. CONCLUSIONS

The interfacial electrical properties of $HfO_2$/GaN and $HfO_2$/AlGaN/GaN including band offsets, interface trap densities and their distribution within the band gap were reported. Optimized $HfO_2$-GaN capacitors show minimal hysteresis in C-V and practically no shift in $V_{FB}$ with dielectric thickness, indicative of low density of fixed bulk and interfacial charges making e-beam evaporated $HfO_2$ an attractive alternative to the more commonly used $Al_2O_3$ dielectrics. A conduction band offset of 1.9 eV was extracted for GaN/$HfO_2$ interface which also displayed a peak trap density of $6.2\times10^{12}$ $eV^{-1}cm^{-2}$ at 0.48eV below $E_C$, and low reverse and forward leakage current densities dominated by Poole-Frenkel emission from traps and Fowler-Nordheim tunneling respectively. Normally-ON HfO$_2$/AlGaN/GaN MISMHETs show negligible shifts in V$_{th}$ post gate deposition and excellent off-state gate leakage currents. Conductance measurements revealed the presence of slow and fast interface trap states at the AlGaN/GaN interface with peak D$_{it}$ >10$^{13}$ eV$^{-1}$cm$^{-2}$. The HfO$_2$/AlGaN interface also shows a peak D$_{it}$ of 4.4x10$^{13}$ eV$^{-1}$cm$^{-2}$, displaying minimal change with V$_G$, suggestive of Fermi level pinning. These results are expected to inform further process development strategies, such as appropriate pre-treatment schemes to de-pin E$_F$ and reduce D$_{it}$ further and thus improve device performance of HfO$_2$ based gate stacks for GaN transistors.

ACKNOWLEDGMENT

The authors would like to acknowledge the National Nano Fabrication Centre (NNFC) and the Micro and Nano Characterization Facility (MNCF) at the Centre for Nano Science and Engineering for access to fabrication and characterization facilities.